\documentclass[ %
 onecolumn, 
 11pt,
nofootinbib,
 amsmath,amssymb,
 aps,
prd,
superscriptaddress
]{revtex4-2}

\usepackage{graphicx}
\usepackage{dcolumn}

\usepackage{lineno}
\usepackage{acro}
\usepackage{booktabs}    
\usepackage{siunitx}
\usepackage{adjustbox}
\usepackage{bm}
\usepackage{multirow}
\DeclareAcronym{LVK}{
  short = LVK ,
  long = LIGO/Virgo/KAGRA ,
  short-plural =  ,
}
\DeclareAcronym{DESI}{
  short = DESI ,
  long = Dark Energy Spectroscopic Instrument ,
  short-plural =  ,
}
\DeclareAcronym{DR2}{
  short = DR2 ,
  long = Data Release 2 ,
  short-plural =  ,
}
\DeclareAcronym{BAO}{
  short = BAO ,
  long = baryon acoustic oscillations ,
  short-plural =  ,
}
\DeclareAcronym{BOSS}{
  short = BOSS ,
  long = Baryon Acoustic Oscillation Survey ,
  short-plural =  ,
}
\DeclareAcronym{6dFGS}{
  short = 6dFGS ,
  long = 6dF Galaxy Survey ,
  short-plural =  ,
}
\DeclareAcronym{SDSS-MGS}{
  short = SDSS-MGS ,
  long = Sloan Digital Sky Survey ``main galaxy sample'' ,
  short-plural =  ,
}
\DeclareAcronym{CMB}{
  short = CMB ,
  long = cosmic microwave background ,
  short-plural =  ,
}
\DeclareAcronym{ACT}{
  short = ACT ,
  long = Atacama Cosmology Telescope ,
  short-plural =  ,
}
\DeclareAcronym{SPT}{
  short = SPT-3G ,
  long = South Pole Telescope ,
  short-plural =  ,
}
\DeclareAcronym{S4}{
  short = S4 ,
  long = CMB Stage-IV ,
  short-plural =  ,
}
\DeclareAcronym{LiteBIRD}{
  short = LiteBIRD ,
  long = Lite satellite for the study of  B-mode polarization and Inflation from cosmic background Radiation Detection ,
  short-plural =  ,
}
\DeclareAcronym{PTA}{
  short = PTA ,
  long = pulsar timing array ,
  short-plural = s ,
}
\DeclareAcronym{CSST}{
  short = CSST ,
  long = China Space Station Telescope ,
  short-plural =  ,
}
\DeclareAcronym{SIGW}{
  short = SIGW ,
  long = scalar-induced gravitational wave ,
  short-plural = s ,
}
\DeclareAcronym{GWB}{
  short = GWB ,
  long = gravitational-wave background ,
  short-plural = s ,
}
\DeclareAcronym{PBH}{
  short = PBH ,
  long = primordial black hole ,
  short-plural = s ,
}
\DeclareAcronym{BHB}{
  short = BHBs ,
  long = black hole binaries ,
  short-plural =  ,
}

\DeclareAcronym{NG15}{
  short = NANOGrav 15yr ,
  long = North American Nanohertz Observatory for Gravitational Waves 15-year ,
  short-plural =  ,
}
\DeclareAcronym{MCMC}{
  short = MCMC ,
  long = Markov-Chain Monte-Carlo ,
  short-plural =  ,
}
\DeclareAcronym{CLASS}{
  short = \texttt{CLASS} ,
  long = Cosmic Linear Anisotropy Solving System ,
  short-plural =  ,
}
\DeclareAcronym{SKA}{
  short = SKA ,
  long = Square Kilometre Array ,
  short-plural =  ,
}
\DeclareAcronym{IR}{
  short = IR ,
  long = infrared ,
  short-plural =  ,
}

\begin{document}

\title{Reassessing the SIGW Interpretation of PTA Signal: The Role of Third-Order Gravitational Waves and Implications for the PBH Overproduction}

\author{Zhi-Chao Zhao
}
\affiliation{%
Department of Applied Physics, College of Science, China Agricultural University, 17 Qinghua East Road, Haidian District, Beijing 100083, China}%
\author{Sai Wang}%
 \thanks{Contact author}%
 \email{wangsai@hznu.edu.cn}
\affiliation{
School of Physics, Hangzhou Normal University, No.2318 Yuhangtang Road, Yuhang District, Hangzhou 311121, China}
\author{Qing-Hua Zhu}%
\affiliation{%
School of Physics, Chongqing University, Chongqing 401331, China}%
\author{Xin Zhang}%
\affiliation{%
Liaoning Key Laboratory of Cosmology and Astrophysics, College of Sciences, Northeastern University, Shenyang 110819, China}%
\affiliation{%
National Frontiers Science Center for Industrial Intelligence and Systems Optimization, Northeastern University, Shenyang 110819, China}%
\affiliation{%
MOE Key Laboratory of Data Analytics and Optimization for Smart Industry, Northeastern University, Shenyang 110819, China}%

\begin{abstract}

In light of recent interpretations attributing pulsar timing array (PTA) signal to second-order gravitational waves induced by linear cosmological curvature perturbations in the early universe, the overproduction of primordial black holes (PBHs) poses a theoretical tension. In this work, we address this issue through extending such a scalar-induced gravitational wave (SIGW) framework to include third-order gravitational waves, which allow for a substantial enhancement in the spectral amplitude of SIGWs. Analyzing a combined dataset from cosmic microwave background and baryon acoustic oscillations, we derive cosmological constraints on the physical energy-density fraction of cosmological gravitational waves. Further incorporating PTA data, we obtain constraints on the spectral amplitude and peak frequency of SIGWs. Our results indicate that the parameter region favored by the data combination can to some extent alleviate the PBH overproduction problem, thereby supporting the theoretical consistency of our model. Furthermore, we demonstrate the robustness of our SIGW interpretation for the PTA signal by extending the analysis to include a gravitational wave background from supermassive black hole binaries. These findings are poised for further scrutiny with future high-precision observations.

\end{abstract}

\maketitle
\flushbottom

\section{Introduction}\label{sec:intro}

Evidence for a nanohertz \ac{GWB} has recently been reported with high significance by multiple \acp{PTA}, with the inferred signal consistent with the expected spatial correlations of an isotropic background~\cite{NANOGrav:2023gor,EPTA:2023fyk,Reardon:2023gzh,Xu:2023wog}.
While an astrophysical origin from supermassive \ac{BHB} remains a well-motivated explanation, the \ac{PTA} band is also sensitive to early-Universe mechanisms that generate gravitational radiation long before recombination~\cite{Antoniadis:2023xlr,NANOGrav:2023hfp,NANOGrav:2023hvm,Bi:2023tib}.
Establishing or ruling out a cosmological origin is therefore of broad interest. It would open a window onto primordial physics at scales far smaller than those directly probed by the \ac{CMB}.
For recent broad reviews on cosmological stochastic gravitational-wave backgrounds and source classes (e.g., see Ref.~\cite{Bian:2025ifp}).

Among cosmological interpretations, the \acp{SIGW} provide a particularly predictive framework.
In this scenario, enhanced primordial linear curvature perturbations on small scales source tensor perturbations at nonlinear order when the corresponding modes reenter the horizon, yielding a stochastic background whose spectrum is calculable once the small-scale curvature power spectrum, denoted as ${\cal P}_\zeta(k)$, is specified~\cite{Ananda:2006af,Baumann:2007zm,Mollerach:2003nq,Assadullahi:2009jc,Espinosa:2018eve,Kohri:2018awv}.
A central feature of the \ac{SIGW} framework is that the same enhancement of curvature perturbations that boosts the tensor perturbations generically also triggers the formation of \acp{PBH} (e.g., see reviews in Ref.~\cite{Carr:2020gox}).
As a result, \acp{SIGW} and \acp{PBH} are tightly linked probes of the small-scale curvature sector, with the \ac{PTA}-sensitive frequencies mapping to \ac{PBH} masses in the sub-solar regime in many benchmark setups.

This tight connection immediately leads to the {\ac{PBH} overproduction problem}~\cite{NANOGrav:2023hvm}.
In the standard Gaussian treatment, fitting a \ac{PTA}-level \ac{SIGW} signal with the conventional {second-order} induced spectrum typically requires an enhancement of the small-scale curvature power to the order of ${\cal O}(10^{-2}$--$10^{-1})$.
However, \ac{PBH} formation depends exponentially on the variance and, more generally, on the tail of the coarse-grained curvature distribution.
Consequently, parameter values that reproduce the \ac{PTA} signal can predict \ac{PBH} abundances that overshoot observational bounds by several orders of magnitude, rendering a purely Gaussian second-order gravitational-wave interpretation strongly constrained or disfavored quantitatively.
A substantial literature has explored ways to relieve this tension while maintaining a cosmological origin, including engineered primordial non-Gaussianity, nonstandard early-time expansion histories, and additional model-building ingredients that alter the mapping between ${\cal P}_\zeta(k)$, \acp{SIGW}, and \acp{PBH} ~\cite{NANOGrav:2023hvm,Franciolini:2023pbf,Wang:2023ost,Liu:2023ymk,Cai:2023dls,Inomata:2023zup,Inomata:2023drn,Abe:2023yrw,Yi:2023mbm,Zhou:2023itl,Firouzjahi:2023lzg,You:2023rmn,Bari:2023rcw,Figueroa:2023zhu,Harigaya:2023pmw,Balaji:2023ehk,Chen:2024fir,Firouzjahi:2023xke,Choudhury:2023fwk,Choudhury:2023fjs,Domenech:2024drm,Wang:2023sij,Zhu:2023gmx,Zhou:2024yke,Zhou:2025djn,Wu:2024qdb,Choudhury:2024kjj,Jiang:2025ysb,Feng:2024yic,Roshan:2024qnv,Domenech:2026nun,Zeng:2025cer,Yogesh:2025hll}.

In this context, it is crucial to revisit a theoretical assumption that is often implicit in phenomenological analyses: that the induced background is adequately described by the second-order contribution alone.
When the curvature enhancement is sizable, precisely the regime suggested by \ac{PTA} fits, higher-order contributions to the induced tensor sector can become non-negligible \cite{Wang:2023sij,Ning:2025yvj}.
Recent computations indicate that {third-order} \acp{SIGW} can contribute comparably to, or even dominate, the {integrated} energy density for sufficiently large peak amplitudes, thereby modifying the relation between the inferred curvature power spectrum and the associated \ac{PBH} abundance~\cite{Wang:2023sij,Yuan:2019udt,Zhou:2021vcw,Chang:2022nzu,Chang:2023vjk,Zhou:2024ncc}.
Because \ac{PBH} abundances respond exponentially to the required curvature amplitude whereas the \ac{SIGW} spectrum responds more mildly, any mechanism that increases the gravitational-wave yield at fixed ${\cal P}_\zeta$, including higher-order gravitational-wave contributions, has the potential to alleviate overproduction by lowering the curvature amplitude needed to match the \ac{PTA} signal.

A complementary ingredient is the use of external cosmological information.
Independent constraints on the total energy density of cosmological gravitational waves from the \ac{CMB} anisotropies and \ac{BAO} measurements provide an essentially model-agnostic bound on the \ac{SIGW} energy-density fraction spectrum across frequencies above $\sim10^{-15}\,$Hz, and thus act as a late-time anchor for any early-Universe explanation of the \ac{PTA} signal~\cite{Moore:2021ibq,Maggiore:2018sht,Smith:2006nka,Clarke:2020bil,DESI:2025zgx,Wang:2025qpj}.
Moreover, in practice, \ac{SIGW} fits to the \ac{PTA} band may probe only the \ac{IR} tail of the \ac{SIGW} spectrum for peaked curvature power spectra, which can lead to degeneracies between the peak amplitude and peak scale when \ac{PTA} data are used alone.
Cosmological bounds on the total gravitational-wave energy density therefore play a key role in pinning down the underlying small-scale curvature parameters in a way that is consistent across cosmic epochs.
This complementarity has been emphasized in recent joint \ac{CMB}, \ac{BAO}, and \ac{PTA} analyses of cosmological gravitational-wave scenarios~\cite{Cabass:2015jwe,Liu:2015psa,Vagnozzi:2023lwo,Bringmann:2023opz,Wang:2023sij,Zhu:2023gmx,Tagliazucchi:2023dai,Zhou:2024yke,Zhou:2025djn,Wu:2024qdb,Wu:2025gwt}.

Motivated by the considerations above, we perform a unified Bayesian study of the \ac{SIGW} interpretation of the nanohertz \ac{PTA} signal in a framework that treats \acp{SIGW} consistently up to {third order}.
For definiteness and transparency, we parameterize the small-scale curvature sector with a sharply peaked primordial spectrum characterized by an amplitude $A_\zeta$ and a characteristic scale (equivalently, a peak frequency $f_\ast$), and compute the resulting present-day \acp{SIGW} including both second- and third-order contributions.
We first derive constraints on the integrated energy-density fraction of cosmological gravitational waves from a joint analysis {\color{black}of \ac{CMB} and \ac{BAO} data~\cite{Carron:2022eyg,SPT-3G:2024atg,SPT-3G:2025bzu,ACT:2025qjh,AtacamaCosmologyTelescope:2025blo,DESI:2025zgx}}, and then incorporate \ac{PTA} information, focusing on the \ac{NG15} data release \cite{NANOGrav:2023gor}, to constrain the \ac{SIGW} spectrum parameters.
We further test robustness by extending the inference to include an additional \ac{GWB} component from \ac{BHB}.
Within this combined framework, we identify regions of parameter space preferred by the data in which the \ac{PTA} signal can be reproduced while the implied \ac{PBH} abundance remains subdominant, thereby {significantly alleviating} the \ac{PBH} overproduction tension in Gaussian \ac{SIGW} scenarios. We also provide prospective forecasts for joint constraints that combine \ac{PTA} data with next-generation \ac{CMB} and \ac{BAO} observations~\cite{LiteBIRD:2022cnt,CMB-S4:2016ple,Gong:2019yxt,Miao:2023umi}.

The remainder of the paper is organized as follows.
In Sec.~\ref{sec:edsgw}, we summarize the \ac{SIGW} formalism up to third order and define the gravitational-wave energy-density observables used in the analysis.
In Sec.~\ref{sec:jcigw}, we present our Bayesian methodology and the joint constraints from \ac{CMB}, \ac{BAO}, and \ac{PTA} datasets (with and without considering \ac{BHB}), and we report prospective joint constraints from \ac{PTA} together with next-generation \ac{CMB} and \ac{BAO} mock data~\cite{LiteBIRD:2022cnt,CMB-S4:2016ple,Gong:2019yxt,Miao:2023umi}. 
In Sec.~\ref{sec:pbhop}, we translate the inferred curvature parameters into \ac{PBH} abundances and discuss the implications for the overproduction problem.
We conclude in Sec.~\ref{sec:summa}.

\section{Theoretical perspective of SIGWs}\label{sec:edsgw}

In this section, we briefly summarize the theoretical results of \acp{SIGW} that will be constrained with observational data in the next section.

\subsection{Nonlinear cosmological perturbations}

The energy-density fraction spectrum of \acp{SIGW} is a key physical quantity that connects primordial curvature perturbations with gravitational-wave observations \cite{Kohri:2018awv}. In previous studies \cite{Ananda:2006af,Baumann:2007zm,Mollerach:2003nq,Assadullahi:2009jc,Espinosa:2018eve,Kohri:2018awv}, this spectrum has typically been derived from second-order tensor perturbations, denoted as $h_{ij}^{(2)}$, which are generated by the quadratic coupling of linear cosmological scalar perturbations. However, when the amplitude of these perturbations is sufficiently large, it becomes theoretically necessary to account for tensor perturbations produced by higher-order couplings of these perturbations. The primary next-order correction leads to third-order tensor perturbations, denoted as $h_{ij}^{(3)}$, arising from the cubic couplings of these perturbations \cite{Zhou:2021vcw,Wang:2023sij}. In this study, we will employ a theoretical framework for \acp{SIGW} that incorporates both second- and third-order tensor perturbations, in order to analyze the latest \ac{PTA} data and address the overproduction problem of \acp{PBH}. 

Specifically, the second- and third-order tensor perturbations are defined in the perturbed spatially-flat Friedmann-Robertson-Walker metric as~\cite{Wang:2023sij,Zhou:2021vcw}
\begin{eqnarray}
\mathrm{d}s^{2} = a^{2}(\eta)\bigg\{&-&\left(1+2 \phi^{(1)}+ \phi^{(2)}\right) \mathrm{d} \eta^{2} + V_i^{(2)} \mathrm{d} \eta \mathrm{d} x^{i} \nonumber \\ &+& \left[\left(1-2 \psi^{(1)} -\psi^{(2)}\right) \delta_{i j}+\frac{1}{2} h_{i j}^{(2)}+\frac{1}{6} h_{i j}^{(3)}\right]\mathrm{d} x^{i} \mathrm{d} x^{j}\bigg\} \,,
\label{eq:metric}
\end{eqnarray} 
where the superscript $^{(n)}$ denotes perturbations of $n$-th order, $a(\eta)$ represents the scale factor of the universe at a conformal time $\eta$, $\phi$ and $\psi$ stand for scalar perturbations, $V_{i}$ denotes vector perturbations, and $h_{ij}$ represents tensor perturbations.  
Here, all perturbations are calculated in the conformal Newtonian gauge \cite{Ma:1995ey}. 
In this work, we neglect the linear anisotropic stress, i.e., $\psi^{(1)}= \phi^{(1)}$. 
In the early universe after the end of inflation, the linear (comoving) curvature perturbations are given by \cite{Dodelson:2003ft}
\begin{eqnarray}
  \zeta =\frac{3}{2}\psi^{(1)}\Big|_{\eta=0}\,. \label{2.2}
\end{eqnarray}
They are frozen on superhorizon scales, and begin to evolve after horizon reentering. 
The power spectrum of primordial linear curvature perturbations is defined by 
\begin{eqnarray}
  \langle \zeta_{\bm k} \zeta_{ \bar{\bm k}} \rangle =(2\pi)^3\delta(\bm k +\bar{ \bm  k})\mathcal{P}_\zeta(k)\,,\label{2.7}
\end{eqnarray}
where $\bm k$ is the wavevector, and $k$ is the wavenumber. 
In this study, we take $\mathcal{P}_{\zeta}(k)$ to be a monochromatic function of $k$, namely, \cite{Kohri:2018awv, NANOGrav:2023hvm} 
\begin{equation}
\mathcal{P}_{\zeta}(k)=A_\zeta k_{\ast}\delta(k-k_{\ast})\,,
\label{eq:primordial_spectrum}
\end{equation}
where $A_{\zeta}$ is the spectral amplitude, and $k_\ast$ is a characteristic wavenumber. 
Both $A_{\zeta}$ and $k_\ast$ are independent parameters to be inferred in Section~\ref{sec:jcigw}. {We emphasize that the monochromatic spectrum in Eq.~(\ref{eq:primordial_spectrum}) is adopted as a controlled benchmark rather than as a shape-independent assumption. Finite-width primordial spectra, such as log-normal or box-like profiles, are physically well motivated and may lead to quantitatively different posterior constraints on $A_{\zeta}$, $f_\ast$, and $f_{\rm PBH}$~\cite{NANOGrav:2023hvm}. However, extending the present third-order calculation to such spectra is not straightforward. In the monochromatic case, the delta function in $\mathcal{P}_{\zeta}(k)$ greatly simplifies the momentum convolutions entering the third-order kernel. For a finite-width spectrum, this simplification is lost, and the calculation requires a genuinely higher-dimensional convolution over the internal momenta.}

The equations of evolution for the second- and third-order tensor perturbations are determined by Einstein's gravitational field equations. To be specific, they are given by~\cite{Wang:2023sij}
\begin{eqnarray}
 h_{i j}^{(2)''}+2 \mathcal{H}  h_{ i j}^{(2)'}-\Delta h_{i j}^{(2)}&=&-4 \Lambda_{i j}^{l m} \mathcal{S}^{(2)}_{l m}[\psi^{(1)},\psi^{(1)}] \,, \label{eq:eom1}\\
 h_{i j}^{(3)''}+2 \mathcal{H}  h_{ i j}^{(3)'}-\Delta h_{i j}^{(3)}&=&-12 \Lambda_{i j}^{l m} \Big(\mathcal{S}^{(3)}_{l m}[\psi^{(1)},\psi^{(1)},\psi^{(1)}] +\mathcal{S}^{(3)}_{l m}[\psi^{(2)}[\psi^{(1)},\psi^{(1)}],\psi^{(1)}] \nonumber \\& &\qquad\qquad +\mathcal{S}^{(3)}_{l m}[\phi^{(2)}[\psi^{(1)},\psi^{(1)}],\psi^{(1)}]+\mathcal{S}^{(3)}_{l m}[V^{(2)}[\psi^{(1)},\psi^{(1)}],\psi^{(1)}] \nonumber \\& &\qquad\qquad +\mathcal{S}^{(3)}_{l m}[h^{(2)}[\psi^{(1)},\psi^{(1)}],\psi^{(1)}]\Big) \,,\label{eq:eom2}
\end{eqnarray}
where $\mathcal{H}=\mathcal{H}(\eta)$ is the conformal Hubble parameter, $\Delta$ is the Laplacian operator, $\Lambda_{ij}^{lm}$ is the transverse-traceless projection operator, and $\mathcal{S}_{lm}^{(2)}$ and $\mathcal{S}_{lm}^{(3)}$, respectively, stand for the second- and third-order source terms. 
Here, we show that $h_{ij}^{(2)}$ is sourced by quadratic couplings of $\psi^{(1)}$ \cite{Kohri:2018awv,Espinosa:2018eve}, while $h_{ij}^{(3)}$ is sourced by cubic couplings of $\psi^{(1)}$ \cite{Zhou:2021vcw,Wang:2023sij,Yuan:2019udt}. 
The second-order perturbations $\phi^{(2)}$, $\psi^{(2)}$, $V^{(2)}_i$, and $h_{ij}^{(2)}$ are also sourced by quadratic couplings of $\psi^{(1)}$, acting as intermediate sources for $h_{ij}^{(3)}$~\cite{Wang:2023sij,Zhou:2021vcw}. The explicit formulas for $\mathcal{S}_{lm}^{(2)}$ and $\mathcal{S}_{lm}^{(3)}$ are outlined in Ref.~\cite{Wang:2023sij}. 
Here, we disregard the dissipation effect due to neutrino diffusions when decoupling, since this effect is negligible to our study \cite{Yu:2024xmz,Yu:2025cqu,Domenech:2025bvr}.

For the second- and third-order tensor perturbations, respectively, the power spectra are defined by their two-point correlations, i.e., 
\begin{equation}
\langle h_{ij,\bm k}^{(n)} h^{(n),ij}_{ \bar{\bm k}} \rangle = 2(2\pi)^3\delta(\bm k +\bar{ \bm  k})\mathcal{P}_h^{(n)}(k,\eta)\,,
\end{equation}
where $n=2,3$ denotes the order of tensor perturbations, $\bm k$ still denotes the wavevector, and $k$ still denotes the wavenumber. 
Further combining with Eqs.~(\ref{2.2},\ref{2.7},\ref{eq:eom1},\ref{eq:eom2}), we get their formulas, i.e.~\cite{Wang:2023sij,Zhou:2021vcw}
\begin{eqnarray}
\mathcal{P}_{h}^{(2)} &=& \frac{k^3}{4 \pi} 
\left( \frac{2}{3} \right)^4  
\int \frac{d^3p}{|\bm{k}-\bm{p}|^3  |\bm{p}|^3} \mathcal{P}_{\zeta}(|\bm{k}-\bm{p}|)  \mathcal{P}_{\zeta}(p)  \mathcal{K}^{(2)}(\bm{k}, \bm{p}, \eta)\,, \label{eq:Ph1}\\
\mathcal{P}_{h}^{(3)} &=& \frac{k^3}{32\pi^2} 
\left( \frac{2}{3} \right)^6  \int \frac{d^3p d^3q}{|\bm{k}-\bm{p}|^3 |\bm{p}-\bm{q}|^3 |\bm{q}|^3} \mathcal{P}_{\zeta}(|\bm{k}-\bm{p}|) \mathcal{P}_{\zeta}(|\bm{p}-\bm{q}|) \mathcal{P}_{\zeta}(q)  \mathcal{K}^{(3)}(\bm{k}, \bm{p}, \bm{q}, \eta)\,, \nonumber \\ 
&&\label{eq:Ph2}
\end{eqnarray} 
where $\mathcal{K}^{(2)}$ and $\mathcal{K}^{(3)}$, respectively, stand for kernels for the source terms $\mathcal{S}_{lm}^{(2)}$ and $\mathcal{S}_{lm}^{(3)}$, as outlined in Ref.~\cite{Wang:2023sij}. 
Further considering Eqs.~(\ref{eq:primordial_spectrum}), we get  
\begin{eqnarray}
&&\mathcal{P}_{h}^{(2)} \propto A_\zeta^2\,,\\
&&\mathcal{P}_{h}^{(3)} \propto A_\zeta^3\,, 
\end{eqnarray}
which indicates that, compared with the second-order tensor perturbations, the third-order tensor perturbations are expected to be subdominant when $A_{\zeta}$ is smaller than a critical value, and vice versa.

\subsection{SIGW energy-density fraction spectrum}

\begin{figure}
    \centering
    \includegraphics[width=\textwidth]{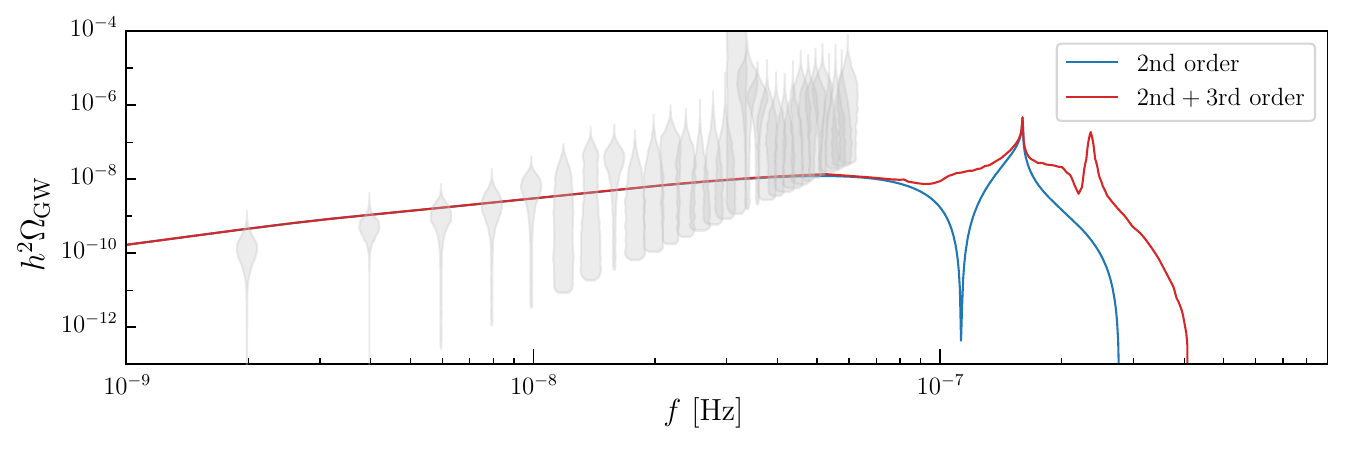}
    \caption{Illustrative figure for the \ac{SIGW} energy-density fraction spectra. The blue solid curve shows the contribution from second-order components only, while the red solid curve includes both second- and third-order components. Both spectra are computed using the same parameter set $(A_\zeta, f_\ast)$, corresponding to the median values in the joint \ac{PTA}, \ac{CMB}, and \ac{BAO} constraints on the \ac{SIGW} spectrum (i.e., the first row of Tab.~\ref{tab:inferenceres}). The grey violins represent the \ac{NG15} dataset \cite{NANOGrav:2023hvm}.}
    \label{fig:illustration}
\end{figure}

Let us introduce one of the most important observables. 
Incorporating both the second- and third-order tensor perturbations, the energy-density fraction spectrum during the epoch of radiation domination is given as~\cite{Wang:2023sij}\footnote{This formula is not the same as that of Ref.~\cite{Wang:2023sij}, as there are typos in the latter. In Ref.~\cite{Wang:2023sij}, extra factors of $A_{\zeta}$ in $\Omega_{\mathrm{GW}}$ should be removed, since they are already included in $\mathcal{P}_{h}^{(2)}$ and $\mathcal{P}_{h}^{(3)}$.}
\begin{equation}
\Omega_{\mathrm{GW}}(k, \eta) = \frac{1}{24} \left( \frac{k}{\mathcal{H}} \right)^2 \left(  \mathcal{P}_{h}^{(2)}(k, \eta) + \frac{1}{9} \mathcal{P}_{h}^{(3)}(k, \eta) \right)\,,
\label{eq:Omega_GW_total}
\end{equation}
where $\mathcal{P}_{h}^{(2)}$ and $\mathcal{P}_{h}^{(3)}$, respectively, are given by Eq.~(\ref{eq:Ph1}) and Eq.~(\ref{eq:Ph2}). 
As the observable, the present-day energy-density fraction spectrum of \acp{SIGW} is given by \cite{Wang:2019kaf}
\begin{equation}
\Omega_{\mathrm{GW},0}(k) = \Omega_{\mathrm{r},0}
     \left(\frac{g_{\ast,\rho}(T)}{g_{\ast,\rho}(T_{\mathrm{eq}})}\right)\left(\frac{g_{\ast,s}(T_{\mathrm{eq}})}{g_{\ast,s}(T)}\right)^{\frac{4}{3}}
     \Omega_{\mathrm{GW}}(k,\eta)\,,
\label{eq:omega_gw_today}
\end{equation}
where $\Omega_{r,0} \simeq 4.2 \times 10^{-5}h^{-2}$ is the present-day energy-density fraction of radiation \cite{Planck:2018vyg}, $h$ is the dimensionless Hubble constant, and both $g_{*,\rho}(T)$ and $g_{*,s}(T)$ are the effective numbers of relativistic degrees of freedom at the cosmic temperature $T$ \cite{Saikawa:2018rcs}. 
The subscript $_{\rm eq}$ denotes quantities at the epoch of matter-radiation equality. 
Here, we present the relation between the frequency $f= k / (2\pi)$ and $T$ as~\cite{Zhao:2022kvz}
\begin{equation}
\frac{f}{\mathrm{nHz}} = 26.5 \left(\frac{T}{\mathrm{GeV}}\right) \left(\frac{g_{\ast,\rho}(T)}{106.75}\right)^{\frac{1}{2}}\left(\frac{g_{\ast,s}(T)}{106.75}\right)^{-\frac{1}{3}} \,.
\label{eq:k_T_relation}
\end{equation}
For the sake of illustration, we depict the energy-density fraction spectra of \acp{SIGW} in Fig.~\ref{fig:illustration}. 
The contribution of third-order tensor perturbations to the \ac{SIGW} spectrum mainly occurs near the peak frequency, i.e., $k \sim k_{\ast}$, due to the resonance amplification, while its contribution in the lower frequency band is not particularly significant. If we fit the \ac{PTA} data using the \ac{IR} tail of the \ac{SIGW} spectrum, analyses based on the \ac{SIGW} spectrum with and without third-order tensor perturbations are expected to yield almost identical constraint results~\cite{Wang:2023sij}. However, due to the significant enhancement of the \ac{SIGW} spectrum near the peak frequency, the total energy-density fraction will be correspondingly increased. We therefore anticipate that cosmological data such as \ac{CMB} and \ac{BAO} will impose tight constraints on this effect~\cite{Clarke:2020bil,Wang:2025qpj}. Based on the above reasons, by combining \ac{CMB}, \ac{BAO}, and \ac{PTA} observational data, we expect to derive new constraints on the \ac{SIGW} spectrum, thereby possibly suppressing both $A_{\zeta}$ and $k_{\ast}$.

\subsection{Total energy-density fraction}

The total physical energy-density fraction of \acp{SIGW} in the present-day universe is defined by an integral of the form
\begin{equation}
\omega_{t} \equiv \omega_{t}^{(2)} + \omega_{t}^{(3)} = \int_{f_{\rm{min}}}^{\infty} h^{2}\Omega_{\rm{GW},0}(2\pi f) \ d\ln f \,, 
\end{equation}
where the lower boundary is $f_{\rm{min}}\simeq3\times10^{-17}$ for the \ac{CMB} \cite{Maggiore:2018sht}. 
Here, the components $\omega_{t}^{(2)}$ and $\omega_{t}^{(3)}$, respectively, represent contributions from the second- and third-order tensor perturbations. 
In addition, we further introduce their ratio of the form  
\begin{equation}
R=\frac{\omega_{t}^{(3)}}{\omega_{t}^{(2)}}\,.
\end{equation}
When they contribute equally to the \acp{SIGW}, namely $R=1$, we find the critical value of $A_{\zeta} \simeq 0.06$, above which the third-order contribution is dominant, otherwise not. 
When discussing the formation of \acp{PBH}, $A_{\zeta}$ is usually required to be on the order of magnitude of ${\cal O}(10^{-2}$--$10^{-1})$ \cite{Carr:2020gox}. This indicates that, compared to second-order tensor perturbations, the contribution of third-order tensor perturbations is also non-negligible.

\section{Performance of \ac{SIGW} interpretation of PTA data}\label{sec:jcigw}

In this section, we first analyze the \ac{CMB} and \ac{BAO} datasets to obtain the posteriors of $\omega_{t}$. Then, we analyze the \ac{PTA} data and incorporate the obtained posteriors of $\omega_{t}$ as a prior of $\omega_{t}$ into the likelihood related to \ac{PTA}. Finally, we derive constraints on the parameters associated with \acp{SIGW} and then the primordial power-spectral amplitude and index using the dataset comprised of \ac{CMB}, \ac{BAO}, and \ac{PTA}.

\subsection{Cosmological analysis and results}\label{ssec:cmbao}

Here, we derive the upper limits on $\omega_{t}$ via analyzing the \ac{CMB} and \ac{BAO} observational and mock datasets, respectively.

\subsubsection{Cosmological model}

The \acp{SIGW} are expected to leave detectable imprints on cosmological probes such as the \ac{CMB} and \ac{BAO}~\cite{Smith:2006nka,Clarke:2020bil}. Under the short-wavelength approximation, \acp{SIGW} can act as an additional energy component in the universe, altering the expansion rate of the early universe and thereby delaying the epoch of matter-radiation equality. This effect will change the size of the acoustic horizon during recombination. On the other hand, fluctuations in the energy density of \acp{SIGW} can affect the time derivative of scalar metric perturbations. This effect will influence the evolution of photon and matter perturbations, leaving detectable signatures in their power spectra. Furthermore, we adopt homogeneous initial conditions for the \ac{SIGW} energy-density fluctuations~\cite{Bucher:1999re,Smith:2006nka,Clarke:2020bil}. Based on the above considerations, we can use the latest \ac{CMB} and \ac{BAO} datasets to constrain the independent parameter $\omega_{t}$.

We study the so-called $w_{0}w_{a}$CDM+$\omega_{t}$ model here. Apart from the independent parameter $\omega_{t}$, this model also has eight additional independent parameters to be inferred. Specifically, $\omega_{b}$ and $\omega_{c}$ represent the present-day physical density fractions of baryons and cold dark matter, respectively. $\theta_{\rm MC}$ is defined as the ratio of the sound horizon to the angular diameter distance at the epoch of decoupling. $\tau$ indicates the Thomson scattering optical depth resulting from reionization. $A_{s}$ and $n_{s}$, respectively, signify the power-spectral amplitude and index of primordial curvature perturbations at the pivot scale $k_{p}=0.05\,{\rm Mpc}^{-1}$. $w_0$ and $w_a$ are used to characterize the dynamical dark energy with the equation of state of the form \cite{Chevallier:2000qy,Linder:2002et}
\begin{equation}
w(a)=w_{0}+w_{a}(1-a)\,,
\end{equation}
where $a$ is the scale factor of the universe. Here, we utilize the \ac{CLASS} \cite{Blas:2011rf} code to generate cosmological models considered in this work.

\subsubsection{Data analysis}

To get observational constraints on $\omega_{t}$ from current cosmological data, the following combination of the \ac{CMB} and \ac{BAO} data is analyzed. For the \ac{CMB}, we utilize the CMB-SPA data combination, as revealed in detail by Tab.~III of Ref.~\cite{SPT-3G:2025bzu}. It incorporates the cutting-edge measurements of the \ac{CMB} temperature anisotropies, polarization, and lensing released by the \emph{Planck} satellite, the \ac{SPT}, and the \ac{ACT}~\cite{Carron:2022eyg,SPT-3G:2024atg,SPT-3G:2025bzu,ACT:2025qjh,AtacamaCosmologyTelescope:2025blo}. Further, we use the recently released \ac{BAO} data from the \ac{DESI} \ac{DR2} \cite{DESI:2025zgx}. 
Here, we consider a uniform prior for $\omega_{t}$, namely $\omega_{t}\in[0,4\times10^{-6}]$. Other priors would not significantly change the leading results of this work. 
In addition, we adopt the \texttt{Cobaya} code \cite{Torrado:2020dgo} for parameter inference following the \ac{MCMC} method.

To get prospective constraints on $\omega_{t}$ from future observations, we consider next-generation experiments that are designed to achieve higher precision. Specifically, we utilize the \ac{LiteBIRD} \cite{LiteBIRD:2022cnt} and the \ac{S4} \cite{CMB-S4:2016ple} ground array for the \ac{CMB} observations, while we use the \ac{CSST} \cite{Gong:2019yxt,Miao:2023umi} for the \ac{BAO} measurements. 
Following the approach of Refs.~\cite{Brinckmann:2018cvx,Audren:2012wb,Wang:2025qpj}, we utilize the mock likelihood of \ac{LiteBIRD} for the \ac{CMB} observations at large angular scales ($2\leq\ell\leq50$), and that of \ac{S4} at small angular scales ($\ell>50$). Further, we utilize the pessimistic precision of \ac{CSST} for the \ac{BAO} measurements, as revealed by Tab.~3 of Ref.~\cite{Miao:2023umi}, indicating conservative constraints on $\omega_{t}$. 
Here, the fiducial model is given by the best-fit parameters inferred from the current cosmological observations, but the fiducial value of $\omega_{t}$ is assumed to be vanishing. 
In addition, we adopt the \texttt{MontePython} code \cite{Brinckmann:2018cvx,Audren:2012wb} for parameter inference following the \ac{MCMC} method.

\subsubsection{Results}

\begin{figure}
    \centering
    \includegraphics[width=0.618\textwidth]{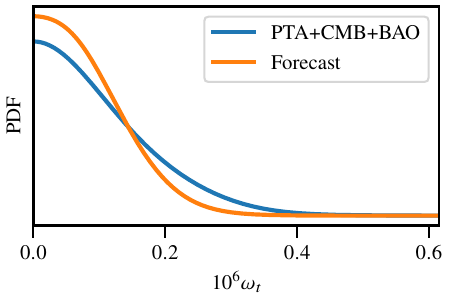}
    \caption{One-dimensional posterior distributions of $\omega_{t}$ obtained via analyzing the CMB-SPA and DESI DR2 observational data (blue) and the LiteBIRD, S4, and CSST mock data (orange).}
    \label{fig:otpost}
\end{figure}

In Fig.~\ref{fig:otpost}, we present the one-dimensional posterior distributions for $\omega_t$, inferred from the observational data (blue curve) and the mock data (orange curve), respectively. In both cases, the 95\% CL upper limits on $\omega_{t}$ are of order $\sim 10^{-7}$. Notably, the posterior derived from the mock data yields tighter constraints on $\omega_t$ than that obtained from the observational data. In the following analysis, we will incorporate these posterior results as priors on $\omega_t$ into the code used to analyze nanohertz gravitational-wave data from \ac{PTA} observations, as demonstrated in the following subsection. This will allow a combined analysis of the latest \ac{CMB}, \ac{BAO}, and \ac{PTA} datasets to infer parameters associated with \acp{SIGW}. Through this approach, we aim to clarify the physical origin of the nanohertz gravitational-wave signal, that is, to determine whether it is cosmological or astrophysical in nature.

\subsection{Methodology incorporating PTA data}\label{ssec:ptada}

We describe the procedure for incorporating \ac{PTA} observations into our joint data analysis. For a given model of the \ac{GWB}, whether cosmological, astrophysical, or a combination thereof, we analyze the \ac{NG15} dataset \cite{NANOGrav:2023gor}, which currently provides the highest statistical confidence among available \ac{PTA} data releases. Following the standard methodology of \ac{PTA} collaborations \cite{NANOGrav:2023hvm}, we perform Bayesian parameter inference using the \texttt{Ceffyl} code \cite{Lamb:2023jls}. The same analysis framework can be directly applied to other \ac{PTA} datasets \cite{EPTA:2023fyk,Reardon:2023gzh,Xu:2023wog} as needed.

Regarding the theoretical interpretation of PTA data, we will consider two distinct models. The first model solely includes \acp{SIGW}, while the second model incorporates not only \acp{SIGW} but also a \ac{GWB} component contributed by supermassive \ac{BHB}. For the former, its energy-density fraction spectrum has already been given in Eq.~(\ref{eq:omega_gw_today}). For the latter, it should further include the energy-density fraction spectrum of the \ac{GWB} generated by supermassive \ac{BHB} that are unresolved individually by \ac{PTA} instruments. This spectrum is typically assumed to follow a power law of the form \cite{NANOGrav:2023hfp,EPTA:2023xxk}
\begin{equation}
\Omega_{\rm{BHB}}(f) = \frac{2\pi^{2}f_{\rm{yr}}^{2}}{3H_{0}^{2}} A_{\rm{BHB}}^{2} \left(\frac{f}{f_{\rm{yr}}}\right)^{5-\gamma_{\rm{BHB}}}\,,\label{eq:omgbhb}
\end{equation}
where $A_{\rm{BHB}}$ and $\gamma_{\rm{BHB}}$, respectively, represent the spectral amplitude and index to be inferred here, $f_{\rm{yr}}$ denotes a pivot frequency corresponding to one year, and $H_{0}$ is the Hubble constant. Therefore, the total spectrum can be expressed as the sum of Eq.~(\ref{eq:omega_gw_today}) and Eq.~(\ref{eq:omgbhb}).

Here, we summarize the model parameters and their prior distributions used in this work. For models involving only \acp{SIGW}, there are two independent parameters, i.e., the amplitude $A_{\zeta}$ and the characteristic frequency $f_{\ast}$, as defined in Eq.~(\ref{eq:primordial_spectrum}). For convenience, we use the corresponding wavenumber $k_{\ast} = 2\pi f_{\ast}$. We adopt uniform (log-flat) priors, namely $\log_{10}A_{\zeta} \in \mathcal{U}(-3, 1)$ and $\log_{10}(f_{*}/\mathrm{Hz}) \in \mathcal{U}(-8, -5)$. For models involving both \acp{SIGW} and \ac{BHB}, we include the two astrophysical parameters $A_{\rm{BHB}}$ and $\gamma_{\rm{BHB}}$ from Eq.~(\ref{eq:omgbhb}), in addition to the primordial parameters above (with identical priors). The priors for $\log_{10}A_{\rm{BHB}}$ and $\gamma_{\rm{BHB}}$ are taken from Ref.~\cite{NANOGrav:2023hvm} and are implemented as a bivariate normal distribution.

When studying \acp{SIGW}, which are of cosmological origin, the corresponding models are constrained not only by the \ac{PTA} observations, but also by the \ac{CMB} and \ac{BAO} data. It is therefore necessary to perform a joint analysis of these complementary datasets.
Specifically, in addition to the cosmological data analyzed in the previous subsection, we incorporate \ac{PTA} observations to jointly constrain the model parameters. To this end, we introduce an informative prior on $\omega_t$ within the \texttt{Ceffyl} code \cite{Lamb:2023jls}, adopting the posterior distributions of $\omega_t$ derived in Fig.~\ref{fig:otpost}. This modification enables a consistent Bayesian inference using the combined dataset.

In order to perform model comparison, we use the Bayes factor, defined as~\cite{kass1995bayes,jeffreys1998theory}
\begin{equation}
    \mathcal{B}_{\alpha\beta}=\frac{p(d\,|\,\mathcal{M}_\alpha)}{p(d\,|\,\mathcal{M}_\beta)}
    =\frac{\int p(d\,|\,{\theta}_\alpha,\mathcal{M}_\alpha)\,\pi({\theta}_\alpha\,|\,\mathcal{M}_\alpha)\,d{\theta}_\alpha}
    {\int p(d\,|\,{\theta}_\beta,\mathcal{M}_\beta)\,\pi({\theta}_\beta\,|\,\mathcal{M}_\beta)\,d{\theta}_\beta}\,,
\end{equation}
where $d$ denotes the observed data, $\mathcal{M}_{\alpha}$ and $\mathcal{M}_{\beta}$ represent two models under consideration, $p(d\,|\,\mathcal{M}_{\alpha})$ is the Bayesian evidence for $\mathcal{M}_{\alpha}$, $p(d\,|\,{\theta}_{\alpha},\mathcal{M}_{\alpha})$ is the likelihood function, and $\pi({\theta}_{\alpha}\,|\,\mathcal{M}_{\alpha})$ denotes the prior distribution for the model parameters ${\theta}_{\alpha}$ in $\mathcal{M}_{\alpha}$. To evaluate the performance of the \ac{SIGW} interpretation of the \ac{PTA} signal, we select $\mathcal{M}_{\beta}$ as the baseline model containing only \ac{BHB}. 
In addition, we adopt the Jeffreys scale for interpreting the strength of evidence based on the Bayes factor, as outlined in Ref.~\cite{jeffreys1998theory}. Specifically, a value of \(\mathcal{B}_{\alpha\beta} > 1\) supports \(\mathcal{M}_{\alpha}\) over \(\mathcal{M}_{\beta}\), with the evidence considered `not worth more than a bare mention' for \(1 < \mathcal{B}_{\alpha\beta} < 10^{1/2}\), substantial for \(10^{1/2} < \mathcal{B}_{\alpha\beta} < 10\), strong for \(10 < \mathcal{B}_{\alpha\beta} < 10^{3/2}\), very strong for \( 10^{3/2}< \mathcal{B}_{\alpha\beta} < 10^{2}\), and decisive for \( \mathcal{B}_{\alpha\beta}>10^{2} \). Conversely, \(\mathcal{B}_{\alpha\beta} < 1\) indicates a preference for \(\mathcal{M}_{\beta}\) over \(\mathcal{M}_{\alpha}\), and the same descriptive scale applies to the reciprocal \(\mathcal{B}_{\beta\alpha}=1/\mathcal{B}_{\alpha\beta}\). When $\mathcal{B}_{\alpha\beta}=1$, both models provide an equally good fit to the data.

\subsection{Results from combined datasets}\label{ssec:resul}

Here, we show the constraints on model parameters inferred from the aforementioned observational and mock datasets, respectively.

\subsubsection{Observational constraints}

\begin{figure}
    \centering
    \includegraphics[width=\textwidth]{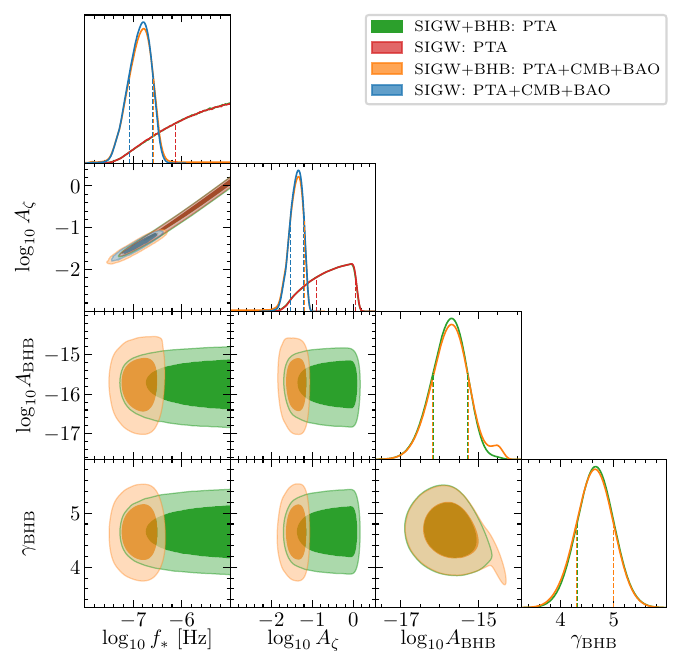}
    \caption{One- and two-dimensional posterior distributions of the independent parameters inferred from the joint data analysis. Dark and light shaded regions, respectively, stand for 68\% and 95\% CL. Dashed vertical lines represent 68\% CL boundaries.}
    \label{fig:posteriorsfin3}
\end{figure}

Based on joint fits to the \ac{PTA}, \ac{CMB}, and \ac{BAO} observations for the models considered in this work, we report the parameter constraints in Tab.~\ref{tab:inferenceres}. This table lists the median values and 68\% credible intervals (or the 95\% credible lower limits where applicable) for the independent model parameters. The corresponding one- and two-dimensional posterior distributions are shown in Fig.~\ref{fig:posteriorsfin3}. Here, the dark and light shaded regions, respectively, stand for 68\% CL and 95\% CL, while the dashed vertical lines represent 68\% CL boundaries.

\begin{table}[h]
\centering
\begin{tabular}{l|l|cccc|c}
\hline
Model & Dataset & $\log_{10} A_{\zeta}$ & $\log_{10} f_{\ast} ~$\rm{[Hz]} & $\mathrm{log}_{10} A_{\mathrm{BHB}}$ & $\gamma_{\mathrm{BHB}}$ & {$\mathcal{B}_{\alpha\beta}$} \\
\hline \cline{2-7}
\multirow{2}{*}{SIGW} & PTA & $> -1.53$ & $> -7.07$ & / & / & $154.50$ \\ \cline{2-7}
 & +CMB+BAO & $-1.38^{+0.15}_{-0.18}$ & $-6.86^{+0.22}_{-0.26}$ & / & / & $28.88$ \\
\hline
\multirow{2}{*}{+BHB} & PTA & $> -1.56$ & $> -7.08$ & $-15.75^{+0.42}_{-0.47}$ & $4.66^{+0.34}_{-0.34}$ & $121.50$ \\ \cline{2-7}
 & +CMB+BAO & $-1.39^{+0.16}_{-0.19}$ & $-6.85^{+0.23}_{-0.27}$ & $-15.72^{+0.46}_{-0.48}$ & $4.64^{+0.35}_{-0.35}$ & $23.77$ \\
\hline\hline
\end{tabular}
  \caption{Median values and 68\% CL uncertainties (or the 95\% credible lower limits where applicable) for the independent model parameters inferred from the joint data analysis. When estimating the Bayes factors, $\mathcal{M}_{\alpha}$ is identified with the model represented by each row of this table, and $\mathcal{M}_{\beta}$ is selected as the baseline model containing only \ac{BHB}.}\label{tab:inferenceres}
\end{table}

We find that when using only the \ac{PTA} data to constrain either the \ac{SIGW} model or the SIGW+BHB model, we obtain only lower limits for the cosmological parameters \(A_{\zeta}\) and \(f_{\ast}\), while their upper bounds reach the prior boundaries and thus cannot be effectively constrained. 
When we plot the \ac{SIGW} spectrum corresponding to the central values of \( A_{\zeta} \) and \( f_{\ast} \) in Fig.~\ref{fig:illustration}, we immediately notice that the peak of this spectrum does not lie within the gravitational-wave frequency band detectable by \acp{PTA}, but is higher by around one order of magnitude. In fact, when fitting the \ac{SIGW} spectrum with \ac{PTA} data, we are essentially fitting its \ac{IR} tail. However, the spectral index of this \ac{IR} tail is universal \cite{Cai:2019cdl,Yuan:2019wwo,Li:2025met}. Therefore, it is difficult to constrain \( A_{\zeta} \), and consequently the range of \( f_{\ast} \), using \ac{PTA} data alone.

In comparison, the combination of \ac{PTA} data with \ac{CMB} and \ac{BAO} observations yields well-constrained posterior distributions for both parameters, with bounds that are fully contained within the prior ranges. This enhanced constraining power stems from the ability of \ac{CMB} and \ac{BAO} data to limit the energy-density fraction of \acp{SIGW}. Joint analysis with \ac{PTA} data thereby significantly reduces the allowable parameter interval for \(A_{\zeta}\), leading to substantially tighter constraints. Notably, the strong limitation placed on the upper bound of \(A_{\zeta}\) further restricts the compatible range of \(f_{\ast}\) due to the positive correlation between the two parameters.

Based on the estimated results for the Bayes factors, we find that the data show a preference for the models containing the \ac{SIGW} component over the model consisting solely of the astrophysical gravitational waves from \ac{BHB}. In particular, using \ac{PTA} data alone yields decisive evidence in favor of the models involving \acp{SIGW}. However, when \ac{PTA} data are combined with the \ac{CMB} and \ac{BAO} observations, the corresponding Bayes factors decrease significantly to a level of strong evidence. This phenomenon reflects the balance between model fit and the added data when assessing the overall evidence. Therefore, based on current data, it cannot be definitively concluded whether the \ac{PTA} signal contains a \ac{SIGW} component. A more conclusive assessment will require future observations with improved precision, e.g., the \acp{PTA} of the \ac{SKA} \cite{SKAOPulsarScienceWorkingGroup:2025oyu,Wang:2022oou,Xiao:2024nmi,Xiao:2025mcg} and the FAST Core Array \cite{ati2024012}.

Moreover, when fitting to the same observational dataset, we find that different models do not significantly shift the posterior distributions of the model parameters, nor do they substantially change the qualitative level of support from the Bayesian evidence, although the Bayes factor decreases somewhat after including the astrophysical gravitational-wave contribution from \ac{BHB}. This indicates strong robustness in our analysis outcomes.

Finally, we reiterate the importance of including the third-order gravitational-wave contribution in interpreting the \ac{PTA} data. While it was found in the literature~\cite{NANOGrav:2023hvm} that \ac{PTA} data alone provide very strong evidence for the \ac{SIGW} interpretation when only the second-order gravitational waves are considered, our present work, which incorporates both second- and third-order contributions, finds that the \ac{PTA} data yield decisive evidence in favor of the \ac{SIGW} scenario. This evidence, however, is reduced to the level of strong evidence once cosmological datasets are included in the analysis.

\subsubsection{Prospective constraints}

\begin{figure}
    \centering
    \includegraphics[width=\textwidth]{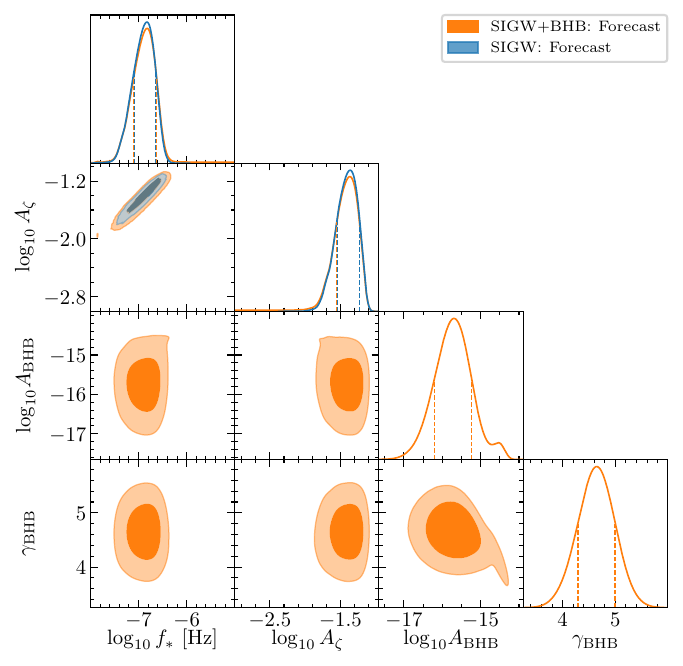}
    \caption{The same as Fig.~\ref{fig:posteriorsfin3}, but we use the mock data of the next-generation \ac{CMB} and \ac{BAO} experiments. }
    \label{fig:posteriorsfin}
\end{figure}

When fitting the future observations to the models considered in this work, we report the prospective parameter constraints in Tab.~\ref{tab:forecast}. This table lists the median values and 68\% credible intervals for the independent model parameters. The corresponding one- and two-dimensional posterior distributions are shown in Fig.~\ref{fig:posteriorsfin}. Here, the dark and light shaded regions, respectively, still stand for 68\% CL and 95\% CL, while the dashed vertical lines represent 68\% CL boundaries.

\begin{table}[h]
\centering
\begin{tabular}{l|cccc}
\hline
Model & $\log_{10} A_{\zeta}$ & $\log_{10} f_{\ast} ~$\rm{[Hz]} & $\mathrm{log}_{10} A_{\mathrm{BHB}}$ & $\gamma_{\mathrm{BHB}}$ \\
\hline
SIGW & $-1.40^{+0.14}_{-0.17}$ & $-6.89^{+0.21}_{-0.25}$ & / & / \\
\hline
+BHB & $-1.41^{+0.15}_{-0.18}$ & $-6.88^{+0.22}_{-0.26}$ & $-15.72^{+0.47}_{-0.48}$ & $4.64^{+0.35}_{-0.36}$ \\
\hline\hline
\end{tabular}
  \caption{The same as Tab.~\ref{tab:inferenceres}, but we use the mock data of the next-generation \ac{CMB} and \ac{BAO} experiments. }\label{tab:forecast}
\end{table}

Comparing Tab.~\ref{tab:forecast} to Tab.~\ref{tab:inferenceres}, we find that the next-generation \ac{CMB} and \ac{BAO} observations do not substantially reduce the allowed parameter space of the models considered here, as further revealed by Fig.~\ref{fig:posteriorsfin}. In fact, as shown in Fig.~\ref{fig:otpost}, these observations do not significantly alter the constraint on the energy-density fraction of the cosmological \ac{GWB}, i.e., $\omega_{t}$ (changing it by less than $\sim10\%$), which in turn is used to constrain the \ac{SIGW}-related parameters. Therefore, we can expect that these data will not lead to a notable tightening of the constraints on those parameters either.

However, it should be emphasized that although our analysis employs the next-generation \ac{CMB} and \ac{BAO} observations, the resulting prospective constraints are not derived from future, more precise \ac{PTA} data, but rather from the current \ac{NG15} dataset. In the upcoming future, observations from the \ac{SKA} and the FAST Core Array are expected to deliver significantly-improved \ac{PTA} measurements, which should help to effectively discriminate between different physical origins of the nanohertz \ac{GWB}. 
Furthermore, we note that 21-cm line observations \cite{Pritchard:2011xb}, which will provide precise measurements of the large-scale structures of the universe \cite{Zhang:2021yof,Wu:2021vfz,Wu:2022dgy,Wu:2022jkf,Pan:2024xoj} and thus place tight constraints on cosmological models, are expected to effectively constrain $\omega_{t}$ as well. We expect that this can further strengthen the constraining power on \ac{SIGW}-related models. Such an investigation, however, lies beyond the scope of the present work and will be explored in detail in future studies.

\section{Implications for \ac{PBH} overproduction problem}\label{sec:pbhop}

When interpreting the \ac{PTA} signal in the framework of \acp{SIGW}, the enhancement of the energy-density spectrum of \acp{SIGW} due to involvement of third-order components results in a suppression of $A_\zeta$.
This amplitude reduction provides a possible resolution to the \ac{PBH} overproduction problem, given the exponential sensitivity of \ac{PBH} abundance to curvature perturbations, i.e., $f_{\mathrm{PBH}} \sim \exp( -{1}/{(2A_{\zeta})} )$ \cite{Green:2020jor,Yoo:2018kvb}.
Here, our analysis demonstrates that the parameter region favored by the combined dataset possibly yields cosmologically acceptable \ac{PBH} abundances, leading to potential reconciliation of the \ac{PBH} overproduction problem.

\subsection{Formulas for \ac{PBH} abundance}

The \ac{PBH} abundance is an integral of the \ac{PBH} mass function $F_{\mathrm{PBH}}(m)$ over the \ac{PBH} mass $m$, namely, 
\begin{equation}
f_{\mathrm{PBH}} = \int F_{\mathrm{PBH}}(m)\ d \ln m\,. 
\end{equation}
Following the theory of critical collapse~\cite{Yokoyama:1998xd,Carr:2016drx} and the Press-Schechter formalism~\cite{Press:1973iz}, we get the \ac{PBH} mass function, i.e.,~\cite{Zhao:2022kvz}
\begin{equation}
F_{\mathrm{PBH}}(m) = \frac{\Omega_{m}}{\Omega_{dm}} \int \tilde{\beta}(m,m_{H}) g(T(m_{H}))\ d \ln m_{H}\,,
\end{equation}
where $\Omega_{m}$ and $\Omega_{dm}$, respectively, stand for the present-day energy-density fraction of non-relativistic matter and dark matter, $m_{H}$ is the mass within the Hubble horizon. 
For the sake of simplicity, we introduce 
\begin{eqnarray}
g(T) &=& \frac{g_{\ast,\rho}(T)}{g_{\ast,\rho}(T_{\mathrm{eq}})}  \frac{g_{\ast,s}(T_{\mathrm{eq}})}{g_{\ast,s}(T)} \frac{T}{T_{\mathrm{eq}}}\,,\\
\tilde{\beta}(m,m_{H}) &=& \frac{\kappa \mu^{\gamma+1}}{\sqrt{2\pi}\gamma\Delta(k)} \mathrm{exp}\left(-\frac{(\delta_{c}+\mu)^{2}}{2\Delta^{2}(k)}\right)\,,
\end{eqnarray}
where $\mu=[m/(\kappa m_{H})]^{1/\gamma}$ has constants $\kappa=3.3$ \cite{Niemeyer:1997mt} and $\gamma=0.36$ \cite{Koike:1995jm, Niemeyer:1999ak, Musco:2004ak, Musco:2008hv, Musco:2012au}, the critical overdensity for the gravitational collapse is $\delta_{c}=0.45$ \cite{Musco:2004ak, Musco:2008hv, Musco:2012au}, and both $g_{\ast,\rho}$ and $g_{\ast,s}$ represent the effective numbers of relativistic degrees of freedom \cite{Saikawa:2018rcs}. 
Here, we can get $T(m_{H})$ via reversing the relation between $m_{H}$ and $T$, namely~\cite{Wang:2019kaf},
\begin{equation}
\frac{m_{H}}{M_{\odot}} = 4.76\times10^{-2} \left(\frac{T}{\mathrm{GeV}}\right)^{-2} \left(\frac{g_{\ast,\rho}(T)}{106.75}\right)^{-\frac{1}{2}}\ ,
\end{equation}
where $M_{\odot}$ denotes the mass of the Sun. 
Moreover, the coarse-grained perturbations during radiation domination are given by~\cite{Young:2014ana,Ando:2018qdb}
\begin{equation}
\Delta^{2}(k) = \frac{16}{81}\int  \left(\frac{q}{k}\right)^{4} w^{2}(\frac{q}{k}) \mathcal{T}^{2}(q,\frac{1}{k}) \mathcal{P}_{\zeta}(q) \  d \ln q \,,
\end{equation}
where $w(y)=\mathrm{exp}(-y^{2}/2)$ is the Gaussian window function, and $\mathcal{T}(q,\tau)=3(\sin x-x \cos x)/x^{3}$ with $x=q\tau/\sqrt{3}$ is the scalar transfer function.

Based on the formulas presented here, for each value of \(k_{\ast} = 2\pi f_{\ast}\), we can derive the corresponding value of \(A_{\zeta}\) for which \(f_{\mathrm{PBH}} = 1\). In Fig.~\ref{fig:tension}, the black dashed curves represent this \(f_{\mathrm{PBH}} = 1\) contour. The parameter regions above the curves correspond to the overproduction of \acp{PBH}.

\subsection{Towards solving \ac{PBH} overproduction problem}

\begin{figure}
    \centering
    \includegraphics[width=\textwidth]{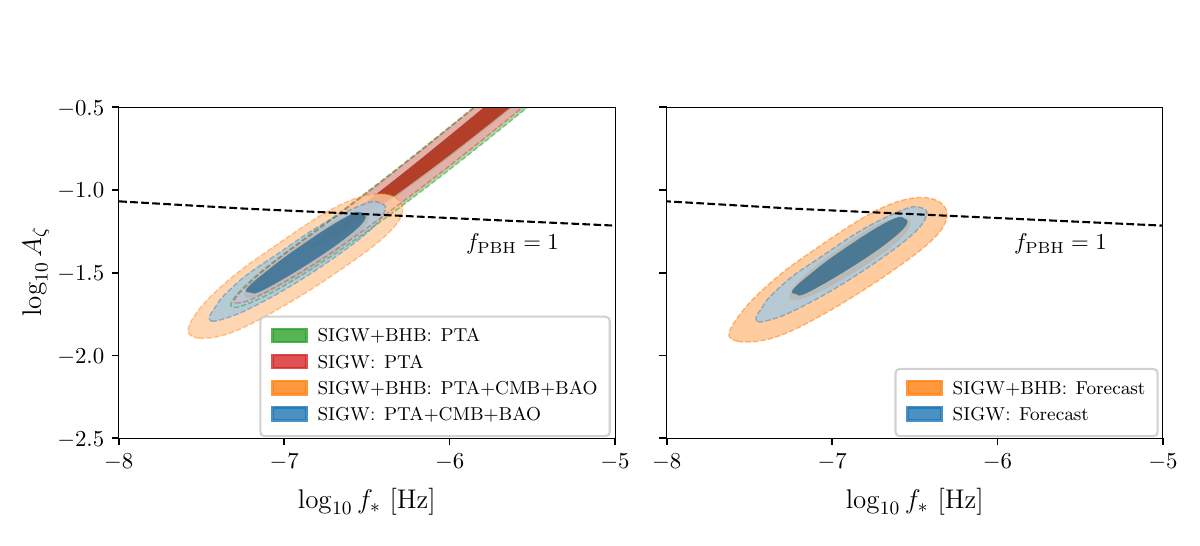}
    \caption{Two-dimensional posterior distributions of $A_{\zeta}$ and $f_{\ast}$ versus the $f_{\mathrm{PBH}}=1$ contour. We depict the left/right panel using the posteriors derived in Fig.~\ref{fig:posteriorsfin3}/ Fig.~\ref{fig:posteriorsfin}. For comparison, we depict the $f_{\mathrm{PBH}}=1$ contour in black dashed curves. }
    \label{fig:tension}
\end{figure}

In Fig.~\ref{fig:tension}, we further compare the two-dimensional posterior distributions of $A_{\zeta}$ and $f_{\ast}$ obtained in Subsection~\ref{ssec:resul} with the $f_{\mathrm{PBH}}=1$ contour. Here, we depict the left panel using the posteriors derived from the current observations, as shown in Fig.~\ref{fig:posteriorsfin3}, while the right panel uses those from the future observations, as shown in Fig.~\ref{fig:posteriorsfin}.

As shown in the left panel of Fig.~\ref{fig:tension}, when only \ac{PTA} data are used, the parameter space favored by the data for the \ac{SIGW} interpretation of the \ac{PTA} signal, regardless of whether the \ac{BHB} background is included, leads to an overproduction of \acp{PBH}. This indicates an internal tension in the theoretical model under this interpretation. This result is consistent with earlier findings in the literature (e.g., Ref.~\cite{NANOGrav:2023hvm}), which reported the same issue when considering only second-order gravitational waves. Our present work, which includes both second- and third-order contributions, confirms that the \ac{PBH} overproduction problem is substantially alleviated but not completely resolved. The underlying reason is that the fit to the \ac{PTA} data in the \ac{SIGW} scenario relies primarily on the universal \ac{IR} tail of the \ac{SIGW} spectrum (illustrated previously in, e.g., Fig.~\ref{fig:illustration}). Because this tail has a universal spectral index, the \ac{PTA} data cannot effectively constrain the upper bound of $A_{\zeta}$. Due to the positive correlation between $A_{\zeta}$ and $f_{\ast}$, the upper bound of $f_{\ast}$ also remains unconstrained. Consequently, the allowed parameter space permits excessively large primordial curvature perturbations, which upon re-entering the Hubble horizon would produce an overabundance of \acp{PBH}.

However, when \ac{CMB} and \ac{BAO} data are incorporated alongside the \ac{PTA} measurements, the allowed parameter space is significantly reduced. Specifically, \(A_{\zeta}\) becomes well-constrained, and consequently, so does \(f_{\ast}\). In this case, the \(f_{\mathrm{PBH}} = 1\) contour only marginally touches the \(68\%\) CL region of the two-dimensional posterior distribution of the model parameters, which substantially alleviates (though does not completely resolve) the \ac{PBH} overproduction problem. This improvement occurs because the \ac{CMB} and \ac{BAO} datasets tightly constrain the total energy-density fraction of the cosmological \acp{SIGW} background, thereby imposing a stringent upper limit on \(A_{\zeta}\) and avoiding parameter regions that would produce an overabundance of \acp{PBH}. Finally, we reiterate that a consistent interpretation of the \ac{PTA} signal within the \ac{SIGW} framework must account for the third-order gravitational-wave contribution, as it becomes comparably important to the second-order contribution when \(A_{\zeta} \sim \mathcal{O}(10^{-2}-10^{-1})\).

As demonstrated in the right panel of Fig.~\ref{fig:tension}, the next-generation \ac{CMB} and \ac{BAO} observations are expected to further lower the upper bound on \(A_{\zeta}\), thereby providing additional mitigation of the \ac{PBH} overproduction issue, though the improvement is relatively modest, with the \(f_{\mathrm{PBH}} = 1\) contour separated from the \(1\sigma\) credible region of the two-dimensional posterior by only about \(1.3\sigma\). It should be noted, however, that while future \ac{CMB} and \ac{BAO} data are used here, the \ac{PTA} dataset remains \ac{NG15} rather than a future \ac{PTA} measurement. Should the precision of \ac{PTA} instruments improve in the future, the quantitative conclusions of this study may change accordingly. Nevertheless, if one aims to investigate the \ac{SIGW} interpretation of the \ac{PTA} signal and the associated \ac{PBH} overproduction problem, relying solely on \ac{PTA} data remains insufficient. Other cosmological probes, such as 21-cm line observations \cite{Pritchard:2011xb}, will continue to be essential. As noted earlier, exploring these possibilities falls beyond the scope of the present work and is left for future study.

\section{Conclusions and discussion}\label{sec:summa}

In this work, we performed a comprehensive Bayesian analysis to reassess the \ac{SIGW} interpretation of the nanohertz \ac{GWB} reported by \acp{PTA}, with particular attention to the associated \ac{PBH} overproduction problem. By consistently incorporating both second- and third-order contributions to the \ac{SIGW} spectrum, and employing a joint dataset comprising \ac{CMB}, \ac{BAO}, and \ac{PTA} (i.e., \ac{NG15}) observations, we derived new constraints on the key parameters of the primordial curvature power spectrum—the amplitude \(A_{\zeta}\) and the characteristic frequency \(f_*\). Our analysis demonstrated that the inclusion of third-order gravitational waves, which become significant for \(A_{\zeta} \sim \mathcal{O}(10^{-2}-10^{-1})\), substantially enhanced the spectral amplitude of \acp{SIGW}. This enhancement allowed the \ac{PTA} signal to be fitted with a lower required curvature perturbation amplitude, thereby providing a crucial mechanism to alleviate the tension with \ac{PBH} overproduction limits. The joint analysis revealed that while \ac{PTA} data alone favored the \ac{SIGW} interpretation decisively, the combination with \ac{CMB} and \ac{BAO} data tightened the constraints on the model parameters significantly, reducing the evidence to the level of "strong" and offering a parameter space where the implied \ac{PBH} abundance could be cosmologically acceptable. 

The integration of cosmological data proved essential for breaking degeneracies inherent in the \ac{PTA}-only analysis. We found that \ac{CMB} and \ac{BAO} observations imposed a stringent upper limit on the total energy-density fraction of cosmological gravitational waves, \(\omega_t\). This external constraint, when combined with \ac{PTA} data, effectively restricted the upper bounds of both \(A_{\zeta}\) and \(f_*\), which were positively correlated. Consequently, the parameter region preferred by the combined data shifted away from the regime that would lead to an overabundance of \acp{PBH}, as defined by the \(f_{\mathrm{PBH}} = 1\) contour. Although the \ac{PBH} overproduction problem was not completely resolved, with the \(f_{\mathrm{PBH}} = 1\) contour remaining close to the \(1\sigma\) credible region of the posterior, the tension was substantially mitigated. This outcome underscored the importance of multi-messenger cosmology, where late-universe anchors from \ac{CMB} and \ac{BAO} were indispensable for pinning down early-universe parameters inferred from gravitational-wave observations.

{
We emphasize that the quantitative conclusions of this work should be understood within the monochromatic benchmark spectrum adopted in Eq.~(\ref{eq:primordial_spectrum}). Finite-width primordial spectra, such as log-normal or box-like profiles, are physically well motivated and may lead to quantitatively different posterior constraints on $A_{\zeta}$, $f_\ast$, and $f_{\rm PBH}$. As discussed in Sec.~\ref{sec:edsgw}, however, extending the present third-order calculation to such spectra is not a straightforward replacement of the primordial spectrum, since the momentum convolutions are no longer collapsed by delta functions, and the third-order kernel becomes a genuinely high-dimensional object. A complete treatment would therefore require a dedicated finite-width third-order convolution calculation, including convergence tests of the high-dimensional kernel interpolation, treatment of integration boundaries and possible resonant regions, and a reassessment of the associated PBH mass function. We leave this important extension to future work, and do not interpret the present result as a shape-independent statement.}

A fully systematic treatment of gravitational waves beyond third order is not yet available in the literature, mainly because the required nonlinear calculations are extremely complicated. We therefore cannot at present establish the full convergence properties of the expansion in Eq.~(15) within a complete resummed framework. Nevertheless, higher-order contributions to the gravitational-wave background are expected to add positive energy density rather than cancel the lower-order terms. Under this expectation, including orders beyond the third would strengthen, rather than weaken, the upper bound on the primordial curvature power spectrum, leading to tighter \ac{PBH} constraints and hence potentially further alleviating the overproduction problem.

A caveat in this study is that the quantitative severity of \ac{PBH} overproduction is not fully model independent. In PTA-motivated \ac{SIGW} interpretations, the inferred \ac{PBH} abundance can shift appreciably when assumptions about primordial non-Gaussianity or the equation of state of the early Universe are changed \cite{Franciolini:2023pbf,Wang:2023ost,Zhu:2023gmx,Liu:2023ymk,Li:2023qua,Li:2023xtl,Chang:2023aba,Firouzjahi:2023xke,Choudhury:2023fwk,DeLuca:2023tun,Iovino:2024sgs,Domenech:2024rks,Liu:2023pau,Pi:2024lsu,Domenech:2020ers}. 
Moreover, this work does not account for evolutionary effects on the \ac{PBH} mass function induced by mergers~\cite{Vaskonen:2019jpv,NANOGrav:2023hvm} or by accretion. In particular, for \acp{PBH} with masses above a few solar masses, accretion can substantially weaken observational constraints~\cite{DeLuca:2020fpg,Ali-Haimoud:2016mbv,NANOGrav:2023hvm}, whereas for sub-solar-mass \acp{PBH} the impact of accretion is expected to be small~\cite{Ali-Haimoud:2016mbv,NANOGrav:2023hvm}. 
Therefore, the alleviation found in this work should be interpreted within the adopted PBH-formation framework, while a sharper conclusion will require further progress on the \ac{PBH} theory side.

Looking ahead, our study highlights several promising directions for future research. Prospective constraints from next-generation \ac{CMB} and \ac{BAO} experiments are expected to provide only a modest further reduction in the allowed parameter space. A more decisive resolution of the \ac{PTA} signal's origin and the \ac{PBH} overproduction issue will likely require significantly improved \ac{PTA} measurements, such as those anticipated from the \ac{SKA} and the FAST Core Array. Furthermore, complementary probes like 21-cm line surveys, which will deliver precise measurements of large-scale structures, hold great potential for independently constraining \(\omega_t\) and thereby strengthening the limits on \ac{SIGW} models. 
In addition, anisotropies of cosmological gravitational-wave backgrounds and their cross-correlations with the \ac{CMB} and large-scale structures can provide complementary diagnostics of source origin and primordial non-Gaussianity. Our work, therefore, not only advances the current understanding of the \ac{SIGW} interpretation and its theoretical consistency, but also provides a clear roadmap for future observational campaigns to definitively test the cosmological origin of the nanohertz \ac{GWB} and its connection to \acp{PBH}.

\acknowledgments

Z.C.Z. is supported by the National Key Research and Development Program of China Grant No. 2021YFC2203001. 
S.W. is supported by the National Natural Science Foundation of China (Grant No. 12533001). 
Q.H.Z. is supported by the National Natural Science Foundation of China (Grant Nos. 12305073, 12547101). 
X.Z. is supported by the National Natural Science Foundation of China (Grants Nos. 12473001, 12575049, 12533001), the National SKA Program of China (Grants Nos. 2022SKA0110200, 2022SKA0110203), the China Manned Space Program (Grant No.CMS-CSST-2025-A02). 
This study is supported by Advanced Computation Center of Hangzhou Normal University.

\bibliographystyle{JHEP}
\bibliography{biblio}

\end{document}